\documentclass[a4paper,12pt]{article}

\usepackage{amsmath}
\usepackage{amsfonts}
\usepackage{amssymb}
\usepackage{bbm}
\usepackage{a4wide}
\def\c #1{{\cal #1}}                             
\def\Dirac{{\raise0.09em\hbox{/}}\kern-0.69em D}
\def\ep{i\epsilon} 

\def\kbar{{\mathchar'26\mkern-9muk}} 		
\def\lesssim{\mathrel{\hbox{\rlap{\hbox{\lower4pt\hbox{$\sim$}}}\hbox{$<$}}}}
\def\sq{\hbox{\rlap{$\sqcap$}$\sqcup$}}         
\def\p{\partial}                                
\def\pprime{{\prime\prime}}                     
\def\tfrac #1#2{\textstyle{\frac{#1}{#2}}} 	

\def\beg{\begin{eg}\rm}                         
\def\eeg{\hfill\sq\end{eg}}                     

\def\Tr{{\rm Tr\,}} 

 \def\cS{{\cal S}} 
\def\cL{{\cal L}}
 
\def\k {\kern-.1em\mathbin{,}\kern-.1em}
\def\hk{\kern.12em\raise-1em\hbox{$\hat{\raise1em\hbox{,}}$}\kern.12em}

\newcounter{eg}                                 
\newtheorem{eg}{Example}[section]
\def\beg{\begin{eg}\rm}                         
\def\eeg{\hfill\sq\end{eg}}                     
        
\newcommand{\initiate}{\setcounter{equation}{0}}        

\def\tX{{\sf X}}

\begin{document}

\title{Spinors on a curved noncommutative space: 
coupling to torsion and the Gross-Neveu model} 
                        \vskip25pt
\author{Maja Buri\'c$^{1}$, \ 
        John Madore$^{2}$ \ 
and Luka Nenadovi\' c$^{1}$\footnote{majab@ipb.ac.rs, madore@th.u-psud.fr, lnenadovic@ipb.ac.rs}
                   \\[15pt]$\strut^{1}$
        University of Belgrade,  Faculty of Physics, P.O. Box 44
                   \\
        SR-11001 Belgrade 
                   \\[5pt]$\strut^{2}$
        Laboratoire de Physique Th\'eorique 
                   \\
        F-91405 Orsay  
       }
\maketitle
\parskip 10pt plus2pt minus2pt
\parindent 0pt
\abstract{We analyse the spinor action on a curved 
noncommutative space, the  so-called  truncated 
Heisenberg algebra, and in particular, the nonminimal 
coupling of spinors to the torsion. We find that 
 dimensional reduction of the 
 Dirac action gives the noncommutative 
extension of the Gross-Neveu model, the model which is, as 
shown by Vignes-Tourneret, fully renormalisable.}

\setlength{\parskip}{5pt plus2pt minus2pt}

\thispagestyle{empty} 

\initiate

\section{Introduction}

A noncommutative model which attracted much attention 
and initiated a great amount of work
in the past decade  is the Grosse-Wulkenhaar (GW) 
model \cite{Grosse:2003nw,Grosse:2004yu}. It describes a
real scalar field $\phi$ on the 
noncommutative Moyal space evolving in the external 
oscillator potential, in two and four euclidean dimensions,\footnote{As
we do not specify the representation here, we have 
only one product: the one which defines the algebra,
(\ref{alg}).}
\begin{equation}
 \cS_{GW} = \int \,\frac 12\, (\p_\mu \phi) (\p^\mu \phi) 
+\frac{1}{2}\,\Omega^2   \tilde x_\mu \tilde x^\mu \phi^2
+\frac{m^2}{2}\, \phi^2 +\frac \lambda 4 \,\phi^4 .
\end{equation}
The model has a number of exceptionally 
good properties in quantisation which have been established 
and analysed in many papers since 2003, and include 
perturbative renormalisability to all orders and vanishing of the 
$\beta$-function at the self-duality point
\cite{Grosse:2004by,Grosse:2004ik}; the model is likely 
to be perturbatively solvable. There is a considerable progress 
in nonperturbative treatment as well; for 
recent results and developments, see 
\cite{Grosse:2012uv,Grosse:2014lxa,Vinas:2014exa}. 
Athough it was initially  treated  in 
the matrix base, the Grosse-Wulkenhaar model was subsequently 
analysed by the multiscale analysis in the coordinate base
\cite{Gurau:2005qm,Gurau:2005gd,Disertori:2006nq}, and that analysis revealed many 
interesting mathematical properties and enabled generalisations.
Similar, though in many aspects different models 
of  fields  in the external
magnetic potential were proposed even before 
\cite{Langmann:2002cc,Langmann:2003if}
as exactly solvable quantum field theories; one of the most
important properties which these models possess is the 
Langmann-Szabo (LS) duality,
 a new kind of symmetry which is present also in the GW model.

Many attempts have been made to understand the physical reasons
underlying renormalisability of the GW model and to generalise 
it to other physical fields, in particular to spinor and gauge fields.
One way of generalisation is  straightforward: by
 constructing Lagrangians which have the Mehler kernel as 
propagator. In the case of spinors this was 
done successfully in \cite{Grosse:2007jy}:
the proposed spinor action is
\begin{equation}
 \cS_{8} =\int \bar\psi {\cal D}_8 \psi
=\int \bar\psi \big( i\Gamma^\mu \p_\mu
 +\Omega \Gamma^{\mu+4}\tilde x_\mu \big) \psi .           \label{D8}
\end{equation}
This action is defined
 on the space of spinors $\,\psi(x^\mu)$, $\, \mu =1,\dots 4$, which
carry a double-dimensional spinor representation:
 $\, \{\Gamma^k,\Gamma^l\} = 2\delta^{kl}$,
$\,k,l = 1,\dots 8$. The square of the  Dirac operator 
${\cal D}_8 $ gives, up to a constant
coordinate-independent matrix $\Sigma$, exactly the Hamiltonian
of the massless GW model. Consequently, the spinor action (\ref{D8}) is 
 renormalisable.
The other possibility of `taking the square root' of the harmonic
potential was proposed in \cite{VignesTourneret:2006nb,Lakhoua:2007ra}. The
2d action which was discussed,
\begin{equation}
 \cS_{n GN} =\int \bar\psi \big( - i\gamma^\mu \p_\mu
 +\Omega \gamma^{\mu}\tilde x_\mu + \tilde m + \kappa \gamma_5\big) \psi
- \frac{g_A}{4}\,{\cal J}_A^2 ,        \label{nGN}
\end{equation}
is a noncommutative extension of the Gross-Neveu (GN) model,
 \cite{Gross:1974jv}; the ${\cal J}_A $ are the currents bilinear 
in the fermionic field.
Remarkably, this action is renormalisable too;
 the parity breaking $\gamma_5$-term appears as  
counterterm when the fermions are massive.

Generalisation of the GW model to gauge fields has been
 more difficult, and indeed  a construction of a renormalisable
gauge model is still an unsolved problem.
 At the first sight the problem is easy 
to understand. In order to have an oscillator-type 
external potential, and correspondingly
the Mehler kernel propagator, one has to include coordinate-dependent 
terms into the action: but coordinate-dependent
terms  break the gauge symmetry.  This particular problem 
however  in noncommutative geometry can be solved surprisingly simply.  
Namely, the momentum operators $\,p_\mu$ which define 
the differential can and often do
belong to the algebra of coordinates $\c{A}$. 
For example on a space with constant nondegenerate noncommutativity
\begin{equation}
 [x^\mu,x^\nu] = i\kbar J^{\mu\nu} ={\rm const}  ,   \label{heis}
\end{equation}
(of which the Moyal space is a representation), the momenta
are given by 
\begin{equation}
   p_\mu = (i\kbar J)^{-1}_{\mu\nu}\, x^\nu.          \label{p}
\end{equation}
 Then covariant derivatives, or more precisely  covariant momenta
$\,  \tilde \tX_\mu = p_\mu + A_\mu   \, $,
also belong to the  space $\c{A}$ and 
moreover transform covariantly,
in the adjoint representation of the gauge group.
(Here by $A_\mu$ we denote the  potential which corresponds to
the gauge group: it is in the literature usually the 
noncommutative $U(1)$ or $U(N)$.) Thus using $\,\tilde \tX_\mu$  or 
covariant coordinates,
$ \, \tX^\mu = x^\mu + i\kbar J^{\mu\nu}A_\nu \, $,
one can define  gauge invariant actions 
\cite{DuboisViolette:1989at,DuboisViolette:1988ps,DuboisViolette:1989vq,book,Madore:2000en}.
Still, additional physical tools  to construct
 the prescribed generalisation of the Yang-Mills action are needed. 
In \cite{Blaschke:2007vc},
 the oscillator potential was introduced through the ghost sector.
A very promising action 
 was obtained in \cite{Grosse:2007dm,de Goursac:2007gq} as an 
effective action for the $U(1)$
gauge field coupled to the GW scalar, after integration 
of the scalar modes. Though these models can be written in terms of 
the covariant coordinates and possess  the LS duality,
they have difficulties related mainly to the vacuum structure 
 and none has proved  to be renormalisable,
\cite{deGoursac:2008rb,Blaschke:2009aw,Buric:2012bb,Blaschke:2013gha,Martinetti:2013uia}.
Comprehensive recent reviews 
of the gauge models are for example, \cite{Martinetti:2013uia,Blaschke:2010kw}.

Another logic of generalisation  of the Grosse-Wulkenhaar model
was proposed in \cite{Buric:2009ss}: it is based on 
the observation that  the harmonic potential can be seen as the 
scalar curvature of an appropriately defined noncommutative space. 
This geometric interpretation gives a straightforward way
to obtain the action  for various fields: it
 is simply the action on a curved spacetime.
There are however additional details.
Since  two-dimensional space (\ref{heis}) 
can be considered as a contraction of a three-dimensional algebra 
(the `truncated Heisenberg algebra') which has 
finite-dimensional matrix representations, the corresponding 
 action can be understood as  geometric, or geometrically 
consistent regularisation. On the level of geometry, we use
 a kind of Kaluza-Klein (KK) reduction followed by
rescaling or renormalisation of the physical fields.
Apart from reinterpretation of the GW model 
\cite{Buric:2009ss,deGoursac:2010zb}, this approach
gave an interesting gauge model \cite{Buric:2010xs}, with
an improved vacuum structure. The model is however 
relatively complicated as due to the KK reduction it
contains interacting gauge and scalar fields: a report on 
the present status of the calculations will be published elsewhere.
 Attempts to define the fermion action in the geometric framework
 were initially not successfull in the sense that 
 coordinate-dependent terms were absent. As we shall see
the reason was simple: we treated only fermions minimally 
coupled to gravity. But even in  commutative case the minimal
coupling, applied in two dimensions, removes the
 explicit dependence on the connection, \cite{nakahara}.
The solution to this problem is to couple fermions to the torsion
nonminimally: a construction of the corresponding  model 
is the main content of this paper, and as we shall see, 
it gives as  result exactly the noncommutative generalisation of the  
Gross-Neveu model studied in \cite{VignesTourneret:2006nb}.

The plan of the paper is the following: in Section~2 we review briefly
 properties  of the truncated Heisenberg space
necessary for our construction and calculate the torsion.
In Section~3 we introduce the  action for the massive Dirac fermions
and  the nonminimal coupling terms  and reduce these actions to
two dimensions. In Section~4 we discuss our results.

\section{The truncated Heisenberg space}

We will shortly introduce the main geometric
objects which are of relevance here. The truncated Heisenberg algebra is
defined by commutation relations
\begin{eqnarray}
&&
[\mu x,\mu y] = i \epsilon (1-\mu z)   \nonumber\\[4pt]
&& [\mu x, \mu z]  = i\epsilon(\mu y \,\mu z+\mu z\, \mu y)           \label{alg}\\[4pt]
&& [\mu y, \mu z] = - i\epsilon(\mu x \,\mu z+ \mu z\,\mu x) .        \nonumber
\end{eqnarray} 
The $\mu\,$ is a constant of dimension of the inverse length; 
$\epsilon $ is a dimensionless parameter which indicates the
strength of noncommutativity; we denote $\kbar = \epsilon \mu^{-2}$.
For $\epsilon =1$ algebra (\ref{alg}) can be represented by 
$\,n\times n$ matrices for any integer $n$, \cite{Buric:2009ss};
$\epsilon =0 $ is  the commutative `limit'. We usually assume that
 parameters can be taken as independent: $\mu$,  related to some 
relevant length or mass scale (like for example the cosmological constant
in the gravitational case),
and $\epsilon\,$ related purely to noncommutativity; one 
 can alternatively  assume just that $\kbar =  l_{Pl}^2$.

Contraction $\mu \to 0\,$   gives the Heisenberg algebra,
\begin{equation}
\qquad\ 
 [x^\mu, x^\nu] =i\kbar\epsilon^{\,\mu\nu}, 
\qquad \mu,\nu = 1,2,
                                       \label{algH}
\end{equation}
which has only infinite-dimensional representations. 
The relation between the Heisenberg algebra and the
truncated  Heisenberg algebra can be seen in the Fock space 
representation of the former as truncation of infinite matrices 
to the finite ones:  it is a weak limit. In many aspects however
it is consistent to  treat this limit as a reduction to the subspace 
$z=0\,$ of the initial noncommutative space, 
\cite{Buric:2011xb}.

Symmetries of  algebra (\ref{alg}) are rotations in the
$xy$-plane: the generator is 
$\, M =\mu^2 x^2 +\mu^2 y^2 +\mu z $.\footnote{On
$z=0\,$ this generator, interestingly, reduces to 
$\, M\vert =  \mu^2 x^2 +\mu^2 y^2 = i \epsilon( x p_2-yp_1 )$.}
Parity on the other hand is not a symmetry
of the truncated Heisenberg algebra, and 
parity breaking  we shall see remains in the
spinor Lagrangian.

Apart from coordinates one can define 
derivations and $p$-forms. In the approach which we are using the space 
of 1-forms is spanned by frame $\{\theta^\alpha \}$, \cite{book}
\begin{equation}
 [f,\theta^\alpha] =0  .                            \label{[f,]}
\end{equation}
 Dual to $\theta^\alpha$  are derivations $e_\alpha$, 
$\,\theta^\alpha(e_\beta) =\delta^\alpha_\beta$.
The differential $d$ of function $f$ can be defined as  
\begin{equation}
 df =( e_\alpha f) \,\theta^\alpha .             \label{df}
\end{equation}
Derivations $e_\alpha$
are inner in the finite-matrix spaces, generated by 
elements $p_\alpha\in \c{A}$
\begin{equation}
 e_\alpha f = [p_\alpha, f]                    \label{ealpha}
\end{equation}
which we call the momenta.
We shall assume that $e_\alpha$ are always of the form (\ref{ealpha})
and in addition that  $\,p_\alpha$ are antihermitian. Condition
(\ref{[f,]}) enables in fact to introduce consistently the metric
which has, in the frame basis, constant components. In our 
particular geometry this metric is euclidean, 
\begin{equation}
 g_{\alpha\beta} =\delta_{\alpha\beta} ,\qquad \alpha,\beta = 1,2,3 .
\end{equation}
As seen from (\ref{df}) the choice of momenta is equivalent to the 
choice of $\,d$: therefore the differential calculus is neither
uniquely fixed, nor do we have a canonical  choice like the 
de Rham calculus in commutative geometry. We choose
for the truncated Heisenberg space, \cite{Buric:2009ss}
\begin{equation}
 \epsilon p_1 =i \mu^2 y, \qquad \epsilon p_2 =-i\mu^2 x,\qquad 
\epsilon p_3 =i\mu (\mu z- \frac 12),
\end{equation} 
so that on $z=0\,$ the differential  reduces to the
standard differential of the Heisenberg space.

The momentum algebra can be used to
 define the exterior product of 1-forms, and to extend this product
 to 2-forms, 3-forms and so on, \cite{book}.  In the truncated  
Heisenberg geometry we obtain the following relations
\begin{eqnarray}
&& (\theta^1)^2 =0,   \qquad (\theta^2)^2 =0,  \qquad (\theta^3)^2 =0, 
\qquad 
 \{ \theta^1,\theta^2 \} =0,     
 \label{2forms}\\[4pt]&& 
\{ \theta^1,\theta^3 \} = \ep (  \theta^2\theta^3 -\theta^3\theta^2),
\qquad  
\{ \theta^2,\theta^3 \} = \ep  ( \theta^3\theta^1 -\theta^1\theta^3).
\nonumber 
\end{eqnarray}
 From (\ref{2forms}) and associativity of the exterior product follow
 the rules of multiplication of three 1-forms:
\begin{eqnarray}
&&
\theta^1\theta^3\theta^1 =\theta^2\theta^3\theta^2, 
\qquad  \theta^3\theta^1\theta^3 =0 , \qquad 
\theta^3\theta^2\theta^3 =0       ,    \nonumber
 \\[4pt]
&& \theta^1\theta^2\theta^3 = -\theta^2\theta^1\theta^3 =
\theta^3\theta^1\theta^2 = -\theta^3\theta^2\theta^1
= i\,\frac{\epsilon^2 -1}{2\epsilon}\,\theta^2\theta^3\theta^2 ,
 \label{3forms}    \\[2pt]
&&\theta^1\theta^3\theta^2 = -\theta^2\theta^3\theta^1 
= i\,\frac{\epsilon^2 +1}{2\epsilon}\, \theta^2\theta^3\theta^2  . \nonumber
\end{eqnarray}
Obviously, there is only one (linearly independent)
 3-form, which means that the volume element is well defined, 
that is unique. We denote it by $\Theta$ and choose 
\begin{equation}
\Theta = - \frac{i}{2\epsilon} \,\theta^2\theta^3\theta^2 
\end{equation}
in order that $\Theta\,$ reduce to $\,\theta^1\theta^2\theta^3 \,$ in the
commutative limit: this is important for example,
 to  properly identify the Lagrangian.
To find the Lagrangian we need in addition the Hodge-$*$ operation.
One possibility to define it, proposed in \cite{Buric:2010xs}, is
\begin{equation}
*[\theta^1,\theta^2] = 2\theta^3, \qquad 
*[\theta^2,\theta^3]= 2\theta^1 , \qquad 
*[\theta^3,\theta^1]= 2\theta^2 .  \label{Hodge1}
\end{equation} 
We shall use this definition: a discussion of its properties
is given in the Appendix.

On noncommutative spaces
 one can define other differential-geometric quantities like the
 affine connection, torsion and curvature. The connection 1-form 
used in \cite{Buric:2009ss} to define the paralell transport on the
truncated Heisenberg space is given by
\begin{eqnarray}
 && \omega_{12} = -\omega_{21} = (-\frac \mu 2 + 2i\epsilon p_3) \theta^3 = \mu\, 
(\frac 12-2\mu z)\theta^3          \nonumber     \\[4pt]
&& \omega_{13} = -\omega_{31} =\frac \mu 2 \,\theta^2 +2i\epsilon  p_2\theta^3 = 
\frac \mu 2\, \theta^2 +2 \mu^2 x\,\theta^3   \label{Con}\\[4pt]
&& \omega_{23} = -\omega_{32} = -\frac \mu 2 \,\theta^1 - 2i \epsilon p_1\theta^3
= -\frac \mu 2 \, \theta^1 +2\mu^2 y\,\theta^3 . \nonumber
\end{eqnarray} 
It can be shown that this paralell transport 
 preserves lengths, that is connection (\ref{Con}) is metric compatible. 
Having the connection,  the torsion and the curvature
tensors are defined as usual:
\begin{eqnarray}
 && T^\alpha = d\theta^\alpha +\omega^\alpha{}_\beta \theta^\beta ,\qquad
 \Omega^\alpha{}_\beta = d\omega^\alpha{}_\beta 
+\omega^\alpha{}_\gamma \omega^\gamma{}_\beta.
\end{eqnarray}
If we denote
\begin{equation}
 \Omega^\alpha{}_\beta = \frac 12\,
R^\alpha{}_{\beta\gamma\delta}\theta^\gamma \theta^\delta
\end{equation}
then by contractions we obtain
the Ricci tensor and the scalar curvature. In our case,
\begin{equation}
R = \eta^{\beta\delta}  R^\alpha{}_{\beta\alpha\delta}= 
\frac{15\mu^2}{2} - 4 \mu^3 z - 8\mu^4  (x^2 +y^2)  .          \label{scalar}
\end{equation} 
Calculating the torsion 2-form we obtain
\begin{eqnarray}
 &&\ \ \ T^1=-i\,\frac{\epsilon\mu}{2}\, (1 -2\mu z)\,[\theta^1,\theta^3]
\nonumber \\[4pt]
&& \ \ \ T^2=-i\,\frac{\epsilon\mu}{2}\, (1-2\mu z)\,[\theta^2,\theta^3] 
\label{T}  \\[4pt]
&&\  \ \ T^3 =-i\epsilon \mu^2 x\,[\theta^2,\theta^3]+i\epsilon \mu^2 y 
\,[\theta^1,\theta^3],
\nonumber
\end{eqnarray}
and its dual
\begin{eqnarray}
 && *T^1=i\epsilon\mu (1-2\mu z)\,\theta^2\qquad\qquad
\nonumber \\[6pt]
&& *T^2=-i\epsilon\mu (1 -2\mu z)\,\theta^1 \label{T*} \\[6pt]
&& *T^3 =-2i\epsilon \mu^2 x\,\theta^1-2i\epsilon \mu^2 y 
\,\theta^2  .
\nonumber
\end{eqnarray}

\section{Spinors on the truncated Heisenberg space}

Let us recall  briefly the commutative action for the Dirac spinors 
in the external gravitational field, to fix the notation. We have
the euclidean space, 
\begin{equation}
\{ \gamma^\alpha, \gamma^\beta \}  = 2\delta^{\alpha\beta},
\end{equation} 
$\alpha,\beta =1,2,3$,\, so the  $\gamma$-matrices are  hermitian.
The Dirac spinor $\psi(x)$  transforms, under the local 
frame rotations,  in the spinor represenatation:
for an infinitesimal rotation $\,\Lambda^\alpha{}_\beta =\delta^\alpha{}_\beta  +\lambda^\alpha{}_\beta\,$ the representation is given by
$\,  S(\Lambda) = 1 + \frac 14 \lambda_{\alpha\beta}\gamma^\alpha \gamma^\beta $.
The covariant derivative is therefore
\begin{equation}
 D \psi = d\psi +\frac 14 \,\omega^{\delta}{}_{\gamma} 
\gamma_\delta \gamma^\gamma \psi = (D_\alpha \psi)\theta^\alpha ,
\end{equation} 
that is,
\begin{equation}
 D_\alpha \psi = e_\alpha \psi +\Gamma_\alpha \psi, \qquad
 D_\alpha \bar \psi = e_\alpha \psi -\bar\psi\Gamma_\alpha , \qquad
 \Gamma_\alpha =\frac 14 \,\omega^\delta{}_{\alpha\beta}\gamma_\delta\gamma^\beta .
\end{equation} 
Since the group generators 
are hermitian,  $\bar\psi = \psi^\dagger$.
The Dirac operator,  $ \,\Dirac = \gamma^\alpha D_\alpha  $
defines the action 
\begin{equation}
{S} =\int \sqrt g\, \bar \psi \,(i\Dirac -m) \psi    .       \label{spinor}
\end{equation} 
It can be seen easily by partial integration that 
(\ref{spinor}) is  real only if the torsion vanishes, more precisely if
\begin{equation}
 \omega^\alpha{}_{\alpha\gamma} =\frac{1}{\sqrt g}\p_\mu (e^\mu_\gamma\, \sqrt g) .
\end{equation} 
If not one defines the spinor action by symmetrisation,
\begin{equation}
 {\cal S} = \frac 12\, (S+S^*).                     \label{symmetr}
\end{equation}

Action (\ref{spinor}) can be rewritten in the language of 
forms, \cite{Aschieri:2011ng}. 
If we introduce a matrix-valued 1-form
$\,V=\theta^\alpha\gamma_\alpha$, in $d$ dimensions we have
 \begin{equation}
\int \Tr  ( D\psi) \bar\psi\,
V V\dots V\gamma_5 =-i(d-1)!\int \Theta \,\bar\psi\gamma^\alpha (D_\alpha \psi),                                     \label{5a}
\end{equation}
where  $\Tr$  is the trace in  $\gamma $-matrices; 
 the product $\,V V\dots V $
contains $(d-1)$ factors and thus the volume 
$d$-form  $\Theta$ appears under the integral. In  commutative
case all 1-forms anticommute so the hermitian part of
(\ref{5a}) is
\begin{equation}
{\cal S}_{kin} =\frac 12 \int \Tr \left((D\psi) \bar\psi
 - \psi (D\bar\psi)\right)V V\dots V\gamma_5 .               \label{5}
\end{equation}
Similarly, the mass term  can be written as
\begin{equation}
 m\int \Tr \psi\bar\psi V V V\dots V\gamma_5 =
-id!\int \Theta\, m\bar\psi\psi,
 \label{6}
\end{equation}
where now the product of 1-forms $\,V V V\dots V $ has 
 $d$ factors.

Since we wish to construct a Dirac spinor on three-dimensional 
space and then reduce  to  two dimensions we need both 
representations. In two and three spacetime dimensions the
 irreducible spinor representations are  two-dimensional.
In 2d a natural choice for the $\gamma$-matrices are Pauli matrices
\begin{eqnarray}
&&  \gamma_1 =\sigma_1
,\quad
 \gamma_2 =\sigma_2
.                \label{12}
\end{eqnarray} 
From $\gamma_1$ and $\gamma_2$ we obtain the  $\gamma_5$-matrix
(which we denote  by $\gamma_3$),
\begin{equation}
 \gamma_3=-i\gamma_1\gamma_2
=\sigma_3  ,                                      \label{3}
\end{equation}
it is the chirality operator in two dimensions.
The representation is, up to unitary equivalence, unique.
In three dimensions the Pauli matrices $\,\gamma_\alpha=\sigma_\alpha$,
$\alpha = 1,2,3\,$ also give  a representation. For the $\gamma_5$ we have
\begin{equation}
 \gamma_4=-i\gamma_1\gamma_2\gamma_3=1 .
\end{equation}
The other, inequivalent representation is 
$\tilde\gamma_1 =\sigma_1$, 
$\tilde\gamma_2 =\sigma_2$,
 $\tilde\gamma_3 = -\sigma_3$, and yields $\tilde \gamma_4 = -1$.
Thus on a three-dimensional  space the spinor action 
is the sum of terms
\begin{equation}
 \cS_{kin}=\frac 14\int \Tr \Big( (D\psi)\bar\psi -\psi(D\bar\psi)\Big) VV,
\qquad
\cS_{mass} =\frac i6 \int \Tr \psi\bar\psi \,V VV.
\end{equation}

Let us  construct the  Dirac action on the truncated Heisenberg
space. For simplicity we first calculate
\begin{equation}
 S^*_{kin}=-\frac 12  \int \Tr \psi(D \bar{\psi})\, V V 
=-\frac 12 \int \Tr \, \Xi_\alpha\gamma_\beta
\gamma_\gamma\theta^\alpha\theta^\beta\theta^\gamma,
\end{equation}
 and then symmetrise; we introduce
\begin{equation}
 \Xi_\alpha = \psi (D_\alpha \bar\psi) =\psi\, \big(
 (e_\alpha\bar\psi) -\bar\psi\Gamma_\alpha\big) .
\end{equation}
Using
\begin{eqnarray}
 && \Xi_1 =\psi\, \big( (e_1\bar\psi) 
+\frac{i\mu}{4} \, \bar\psi\gamma_1\big), \qquad \Xi_2 = \psi\, \big( (e_2\bar\psi) 
+\frac{i\mu}{4}\, \bar\psi\gamma_2\big)   ,         \\[4pt]
&&\Xi_3 = \psi\, \big( (e_3\bar\psi) 
-\frac{i\mu}{4}\, \bar\psi\gamma_3
 +i\mu^2 \bar\psi(x\gamma_2 -y\gamma_1+z\gamma_3)\big)     \nonumber
\end{eqnarray}
and the algebra of 1-forms  we obtain
\begin{eqnarray}
&&  S^*_{kin} = \frac12\int\Theta\, \Tr \left( i\,\Xi_1\gamma_1 + i\,\Xi_2\gamma_2 
+i(1-\epsilon^2)\, \Xi_3\gamma_3- \epsilon\, \Xi_1\gamma_2 +\epsilon\, \Xi_2\gamma_1
\right)   \nonumber  \\[4pt]
&&\phantom{ S^*_{kin} }
= \frac 12\int\Theta\, \Big(\, i(e_1\bar\psi)\gamma_1\psi +
i(e_2\bar\psi)\gamma_2\psi  + i (1-\epsilon^2) (e_3\bar\psi)\gamma_3\psi
   \label{22}  \\[4pt] 
&& \phantom{S^*_{kin} =\int } 
-\frac{\mu}{4}\,(1+\epsilon^2) \bar\psi\psi +\frac{\mu\epsilon}{2} \,\bar\psi\gamma_3\psi - \mu^2(1-\epsilon^2) \bar\psi z\psi\nonumber \\[4pt] 
&&\phantom{S^*_{kin} =\int } 
-\epsilon (e_1\bar\psi) \gamma_2\psi +  \epsilon (e_2\bar\psi) \gamma_1\psi 
-i\mu^2(1-\epsilon^2) \bar\psi(x\gamma_1+y\gamma_2)\psi  \Big). \nonumber
\end{eqnarray}
Terms in the last line are imaginary. Therefore taking the hermitian part,
when we reduce (\ref{22}) to subspace $\,z=0$,
 $\, e_3\bar\psi =0$, we obtain
\begin{eqnarray}
 && {\cal S}\vert_{kin}
=\frac 12 \int\Theta\, \Big(\, i\bar\psi \gamma^\alpha (e_\alpha \psi ) - i(e_\alpha\bar\psi)\gamma^\alpha\psi 
+\frac 12\,\mu(1+\epsilon^2)\bar\psi\psi -{\mu\epsilon} \,\bar\psi\gamma_3\psi \Big)
\qquad                                 \label{Sdirac}
\end{eqnarray}
from the kinetic part of the 3d action;
 summation is now in $\alpha =1,2$. We can observe that,
as in the scalar case, a part of the connection 
terms after dimensional reduction manifest as  mass.
In a similar way from the 3d mass term (\ref{6})
we have
\begin{eqnarray}
 &&{\cal S}\vert_{mass} =i \,\frac{m}{6}\int \Tr \psi\bar\psi\, V V V =
- m  \int\Theta\, \Big( (1-\frac{\epsilon^2}{3})\bar\psi \psi -\frac{2\epsilon}{3}
\,\bar\psi\gamma_3\psi \Big) ; \qquad       \label{SdiracM}
\end{eqnarray}
 it looks the same before and after  dimensional reduction. In fact
being more precise, the volume element before dimensional reduction is
$\,\Theta^{(3)}$, and after it is  $\, \Theta^{(2)}$.
This is not explicitly stressed in our notation; it is also 
 understood  implicitly that the
 integral over the third direction renormalizes the
field $\psi$ and, in the interacting case, the coupling constants. 

In the absence of noncommutativity, $\epsilon =0$, the mass term reduces to
the  usual one. But  noncommutative algebra (\ref{alg}) is not invariant
under the space inversion: this is reflected in the
$ \, \bar\psi\gamma_3\psi $ terms in (\ref{Sdirac}) and (\ref{SdiracM}).
Therefore the spinors of different chirality  have different masses
in (\ref{SdiracM}): for
$\epsilon =1$ for example $\psi_R  $ is massless  while 
 $m_L = 4m/3$.  The kinetic term also generates
 different masses for spinors of opposite chirality, 
$\, m_{L,R}=\mu(1\pm\epsilon)^2/4$. 

But as  stressed before, the minimal coupling of spinors to 
the curved background does not give  coordinate-dependent terms:
 their contributions are imaginary.
 Therefore, if we wish to introduce the `square root of the harmonic potential',
 we have to include nonminimal interaction with 
the torsion. Various 
interaction terms are possible, \cite{Shapiro:2001rz}; from dimensional analysis and 
invariance arguments it
 follows that they are linear in  torsion and bilinear in spinors. 
Interaction terms in 3d are
\begin{equation}
 S^\prime_{tor} = \int\Tr \psi\,\bar\psi \, T_\alpha \gamma^\alpha V,\qquad
 S^\pprime_{tor}= \int\Tr \psi\,\bar\psi \,(*T_\alpha)\gamma^\alpha V V,
\end{equation}
but it turns out  that they are proportional,
\begin{equation}
 S^\pprime_{tor} =2i S^\prime_{tor}.
\end{equation}
The calculation gives
\begin{eqnarray}
&& S^\prime_{tor}= 2\epsilon \int\Theta \,\bar\psi\Big(
 (\epsilon-\gamma_3) (\mu-2\mu^2 z)+ (\mu^2 x \gamma_2 -\mu^2 y\gamma_1) \Big) \psi
\nonumber  \\[4pt] &&\phantom{S_1=}
-2i\epsilon^2 \int\Theta \,\bar\psi\Big(\mu^2 x\gamma_1 +\mu^2 y\gamma_2  \Big) \psi,
\end{eqnarray}
so we have two independent  terms,
the real and the imaginary  part of $S^\prime_{tor}$.  
After the reduction to $z=0$ we obtain
\begin{eqnarray}
&& {\cal S}\vert_{tor}=\frac{a\mu}{2} \int\Theta \,( \epsilon \,\bar\psi\psi
-\bar\psi\gamma_3\psi )\,
+  \frac{1}{2}\int\Theta \, \bar\psi (a\epsilon_{\alpha\beta} +b\delta_{\alpha\beta}) \mu^2 x^\alpha\gamma^\beta  \psi, \qquad \           \label{Sd+t}
\end{eqnarray}
where $a$ and $b$ are arbitrary real coefficients and
 the summation is,  as in (\ref{Sdirac}), in $\,\alpha =1,2$.

In conclusion, the
 general Lagrangian which describes the 3d Dirac spinors, after 
the reduction to two dimensions, is given by 
\begin{eqnarray}
 && {\cal L} ={\cal L}\vert_{kin}+{\cal L}\vert_{mass} +{\cal L}\vert_{tor} =
   \label{spin}        \\[6pt]  && 
=\frac 12 \, \Big(\, i\bar\psi \gamma^\alpha (e_\alpha \psi ) - i(e_\alpha\bar\psi)\gamma^\alpha\psi 
 \Big) +  \frac{1}{2} \, \bar\psi (a\epsilon_{\alpha\beta} +b\delta_{\alpha\beta})
 \,\mu^2 x^\alpha\gamma^\beta  \psi     \nonumber
\\[4pt]  && \phantom{\  }
- m \Big( (1-\frac{\epsilon^2}{3})\bar\psi \psi -\frac{2\epsilon}{3}
\,\bar\psi\gamma_3\psi \Big)  
+\frac \mu 4\, \Big((1+ 2a\epsilon +\epsilon^2)\bar\psi\psi 
-{2(\epsilon + a)} \,\bar\psi\gamma_3\psi \Big)  . \quad  \nonumber
\end{eqnarray}
Writing (\ref{spin})  in the form
$\,  \, {\cal L} = \bar\psi\,  \Dirac \psi \,$
we find  the corresponding Dirac operator
\begin{equation}
 \Dirac = i\gamma^\alpha e_\alpha - A- B\gamma_3
 +\frac 12  (a\epsilon_{\alpha\beta} +b\delta_{\alpha\beta})
 \,\mu^2 x^\alpha\gamma^\beta ,
\end{equation}
with
\begin{equation}
 A=  \frac 13 \,(3-\epsilon^2) m  - \frac 14\, (1+ 2a\epsilon +\epsilon^2) \mu,\quad
B=-\frac{2\epsilon}{3}\, m +\frac 12\, (a +\epsilon) \mu.     \label{AB}
\end{equation}

\noindent
The square of this operator is
\begin{eqnarray}
&& \Dirac^2 = -e_\alpha e^\alpha - 2 A\gamma^\alpha ie_\alpha 
+\frac{1}{4}\, (a^2+b^2) \mu^4 x_\alpha x^\alpha
\\[2pt] && \phantom{\Dirac^2 =}
+(A^2+B^2) +\big( 2 AB - \mu^2 a -\frac 14\, \mu^2\epsilon(a^2+b^2)\big)\gamma_3
\nonumber \\[2pt] && \phantom{\Dirac^2 =}
- A (a\epsilon^{\alpha\beta} + b\delta^{\alpha\beta})\mu^2 x_\alpha \gamma_\beta
+ \frac 12\, (a\epsilon^{\alpha\beta} + b\delta^{\alpha\beta})\{ ie_\alpha, \mu^2 x_\beta\} .
\nonumber
\end{eqnarray}
The obtained $\ \Dirac^2\,$ is a generalisation of the
usual  Lichnerowicz spinor laplacian: it contains an
additional dependence on the connection coming from  the
interaction with the torsion. In addition, there are terms 
induced by dimensional reduction.

\section{Concluding remarks}

The main objective of  calculations presented in this paper
 was to extend our previous
work and build a consistent geometrical action for the Dirac spinors 
on the truncated Heisenberg space, and  to 
reduce it in the next step to the Heisenberg subspace. As we 
wished to relate  this theory 
eventually  to the Grosse-Wulkenhaar 
action, we wanted to include coordinates explicitly:
therefore we introduced the nonminimal interaction of spinors with
the torsion. 
The result, somewhat unexpected, is indeed
 nice: the action (\ref{spin}) which we found
is in fact equivalent to the noncommutative extension of the
Gross-Neveu action (\ref{nGN}) proposed by
Vignes-Tourneret, which is renormalisable.
Renormalisability of the noncommutative GN model does not
trivially reduce to renormalisability of the GW model
 since one is not a simple square  of the other, 
\cite{VignesTourneret:2006nb}.

To effect the mentioned equivalence we need in fact only one of
the interaction terms:  we set $\,b=0$. Comparing notations
 we identify noncommutativity $\theta$ of \cite{Lakhoua:2007ra}
as $\theta = -\kbar$. Then  the remaining parameters of the fermion actions
(\ref{nGN}) and (\ref{spin}) are associated  as
\begin{equation}
 \tilde m =A, \qquad \kappa=B,\qquad \Omega = \frac {a\epsilon}{4},
\end{equation}
where $A$ and $B$ are given in (\ref{AB}).

It is further interesting to notice that the harmonic term
 in the Grosse-Wulkenhaar model can be obtained, alternatively,
 as an interaction 
of the scalar field with the torsion. Therefore `geometrisation'
of the mentioned  renormalisable actions can be formulated in
terms of the interaction with torsion solely. Indeed, from (\ref{T}-\ref{T*})
on the truncated Heisenberg space we find
\begin{equation}
(*T_\alpha)\, T^\alpha = T^\alpha \,(*T_\alpha)
  =-2\mu^2 \epsilon^2 \Theta\,\Big(
(1 -2\mu z)^2+2\mu^2(x^2+y^2) -2\epsilon^2(1-\mu z)\Big)  
\end{equation}
so the corresponding interaction Lagrangian with the scalar field $\phi$, 
reduced to 2d, is 
\begin{equation}
\cL_{\phi,tor} =-\xi \mu^2 \epsilon^2 \Big(
2(\mu^2x^2+\mu^2y^2) +1-2\epsilon^2 \Big) \phi^2 ,
\end{equation}
where $\xi/2$ is the coupling constant. As in the case 
of the coupling with curvature \cite{Buric:2009ss}, the interaction
with torsion introduces the harmonic potential and
 modifies the mass. The present result seemingly suggests 
that the torsion in a way has a primary role in the analysed set of models.
Unfortunately it cannot couple to  nonabelian gauge fields
in a gauge invariant way, and on a noncommutative
space all gauge fields including the $U(1)$ are nonabelian.
The question whether the torsion might  improve
properties of the gauge models needs perhaps some further 
clarification, maybe in the view of the dimensional reduction
procedure.

There are other  effects contained in our result which
deserve further investigation: the creation of mass and the parity 
breaking. The fact that the gravitational field 
(seen as curvature, or torsion) manifests itself as inertia, that is
  mass, is intuitively clear.
As an additional possible source of the particle mass we 
have here the dimensional reduction,
but as the extra dimension is not  compact, its volume
 just renormalises the wave function and the couplings.
The parity breaking is also not hard to understand.
Since we start  from a 3d space which is not invariant under
the space inversion, the spinor Lagrangian  does not
have the symmetry either and the property remains 
after reduction to 2d. It is manifested as 
a difference between masses of the 
right and the left components of the spinor field:
we have
\begin{equation}
 m_{R,L} = A\pm B = \frac m3\, (1\mp \epsilon)(3\pm \epsilon)
 -\frac \mu 4 \, (1\mp\epsilon)(1\mp \epsilon\mp 2a)  .
\end{equation}
Interestingly, the effect of the parity breaking can be 
produced solely by coupling to the torsion, which can be seen
in the previous formula by putting $\epsilon =0$, $a\neq 0$. 
Both of the mentioned effects  give interesting possibilities for
the model building in particle physics: they
might provide us with  new variants of the see-saw 
mechanism as well.

\vskip0.5cm\noindent
{\bf Acknowledgement}\  \ 
This work was supported by the Serbian Ministry of Education, Science
and Technological Development Grant ON171031. 

\newpage

\noindent
\begin{large}
{\bf Appendix: Hodge-$*$ and the volume form}
\end{large}
\vskip0.5cm

A part of the algebra of 3-forms  (\ref{3forms}), rewritten as
\begin{equation}
\begin{array}{ll}
[\theta^1,\theta^2]\,\theta^3 = \theta^3
 [\theta^1,\theta^2] = 2(1 -\epsilon^2 )\Theta, \ & 
\\[8pt]
[\theta^2,\theta^3] \,\theta^1 = \theta^1
[\theta^2,\theta^3] = 2\Theta,
& [\theta^3,\theta^1]\,\theta^2 = \theta^2 [\theta^3,\theta^1] = 2\Theta ,
\end{array}                                             \label{3forms1}
\end{equation} 
suggests  definition (\ref{Hodge1}) the $*$-operation.
This definition is further in  accordance with the usual 
convention for the double 
action of the Hodge-$*$, which  in three euclidean dimensions  is
$\ *(*\omega) =\omega\, $ for all $p$-forms $\omega$.
It  looks however as if (\ref{Hodge1}) changes the usual rules related to
the  volume element as for example
\begin{equation}
\left(*\tfrac 12\,[\theta^1,\theta^2]\right)\, 
\tfrac 12\, [\theta^1 ,\theta^2] =(1 -\epsilon^2 )\, \Theta ,\qquad
\left(*\tfrac 12\,[\theta^2,\theta^3]\right)\, 
\tfrac 12\, [\theta^2 ,\theta^3] = \Theta .
\end{equation}
At this point we should recall that, in 
noncommutative case, the commutators
of 1-forms are not natural as a basis in the space of 2-forms:
we should rather use the 
  twisted commutators, $\ \tilde\theta^{\alpha\beta} \equiv P^{\alpha\beta}{}_{\gamma\delta}\theta^\gamma\theta^\delta \, $, as
twisted commutators
 enclose the properties of the noncommutative product.
We have, \cite{Buric:2010xs}
\begin{eqnarray}
&&\theta^1\theta^2 =\tilde \theta^{12} =P^{12}{}_{\gamma\delta}\theta^\gamma\theta^\delta 
=\frac 12\, [\theta^1,\theta^2 ]  \nonumber \\[4pt]
&& \theta^1\theta^3 =\tilde\theta^{13} =P^{13}{}_{\gamma\delta}\theta^\gamma\theta^\delta 
=\frac 12 \,[\theta^1,\theta^3 ] 
+\frac {i\epsilon}{2}\, [ \theta^2,\theta^3 ]  \label{basis} \\[4pt]
&&\theta^2\theta^3=\tilde\theta^{23} = P^{23}{}_{\gamma\delta}\theta^\gamma\theta^\delta 
=\frac 12 \,[\theta^2,\theta^3 ] 
-\frac{ i\epsilon }{2}\,[ \theta^1,\theta^3 ] .  \nonumber
\end{eqnarray}
The main drawback of the basis  $\{ \tilde\theta^{\alpha\beta}\}\,$
 it that is not hermitian. Applied to
 this basis elements  the  Hodge-$*$  gives
\begin{equation}
(*\tilde\theta^{12})\,\theta^{12} = 
(*\tilde\theta^{13})\,\theta^{13} = 
(*\tilde\theta^{23})\,\theta^{23} = \theta^1\theta^2\theta^3
                                    \label{123}
\end{equation} 
as one would expect; also, the order of factors  does not matter. 
From (\ref{123}) we see that  in fact 
the volume 3-form should have been identified as
\begin{equation}
 \widetilde\Theta =\theta^1\theta^2\theta^3 =(1-\epsilon^2)\Theta.  \label{tildeTh}
\end{equation}
However because of non-hermiticity, we have for example
$ \,  (*\tilde\theta^{13})\,\tilde\theta^{12}\neq 0\, $ but
 rather
\begin{equation}
(*\tilde\theta^{13})\,\tilde\theta^{12} +\tilde\theta^{12}\,  (*\tilde\theta^{13})= 0.
\end{equation} 
Analogous relations hold for other components. 
We have not in our calculation redefined the volume 3-form $\,\Theta$
to $\, \widetilde\Theta$, as this redefinition  changes only the
 overall factor
in the action. But formula (\ref{tildeTh})  shows that the limit to the matrix case,
$\epsilon =1$, is not smooth: the space of $p$-forms becames 
a kind of fragmented. In a similar way
the commutative limit, $\epsilon =0$, is singular because in this limit
the momenta $p_\alpha$, (\ref{p}) diverge.  Perhaps a more detailed analysis  
of the tangent and cotangent spaces for $\epsilon=1$ could reveal some interesting 
or characteristic properties of the matrix geometries or of the quantum groups,
of which the truncated Heisenbeg algebra is one, somewhat exotic, example.

\end{document}